\documentclass{article}
\usepackage{amsmath,amsthm}
\usepackage{amssymb,latexsym}
\usepackage[mathscr]{eucal}
\usepackage{mathrsfs}
\usepackage{setspace}
\usepackage{graphics}
\usepackage{array}

\newcommand{\lp}{\left(}
\newcommand{\rp}{\right)}
\newcommand{\lb}{\left[}
\newcommand{\rb}{\right]}

\newcommand{\beq}{\begin{equation}}
\newcommand{\eq}{\end{equation}}

\begin{document}

\title{\textbf{Effects of Stellar Rotation in Parker's Hydrodynamic Stellar Wind Model: How Protostars and Strong Rotators Lose their Angular Momentum Fast }}         % Enter your title between curly braces
\author{B. K. Shivamoggi\\
University of Central Florida\\
Orlando, FL 32816-1364, USA\\
}        % Enter your name between curly braces
\date{}          % Enter your date or \today between curly braces
\maketitle

\noindent \large{\bf Abstract} \\ \\

The effects of the stellar rotation and the consequent azimuthal stellar wind flow in Parker's \cite{Par} hydrodynamic stellar wind model are discussed. Of special interest is the emergence of a whole new \textit{hydrodynamic} physics via a new critical point in the stellar wind flow, which supersedes the critical point in Parker's \cite{Par} hydrodynamic model. The effect of the stellar rotation is shown to cause the new critical point to occur lower in the corona, so the stellar wind experiences a stronger \textit{afterburner} (as in an aircraft jet engine) action in the corona. For strong rotators, the new critical point is shown to occur at a fixed location for a given star, determined only by the basic stellar parameters like the mass $M$ and the angular velocity $\Omega_\ast$, the variations in the stellar wind environment notwithstanding. The stellar rotation leads to stronger density fall-off and enhanced acceleration of the stellar wind at large distances from the star - this effect materializes even close to the star, for strong rotators. The stellar rotation causes the physical throat section of the effective \textit{de Laval} nozzle associated with the stellar wind flow to become narrower, indicative of an enhanced flow acceleration. Thus, the stellar rotation leads to tenuous and faster stellar wind flows without change in the mass flux, and hence provides an efficient physical mechanism for protostars and strong rotators to lose their angular momentum quickly.

\vspace{.20in}

\newpage

\noindent\large\textbf{1. Introduction}\\

Stellar wind is an interplanetary continuous outflow of hot plasma from a star (the solar wind\footnote{The structure of the solar wind is observed by SOHO (Cho et al. \cite{cho}) to change from the solar maximum (when it is latitudinally uniform) to solar minimum period (when it is bimodal - faster in the poles to slower along the equator).} typically emanates from the coronal tides (Sakao et al\cite{sak})) and an associated remnant of the stellar magnetic field that pervades the space surrounding the star (like the heliosphere formed by the solar wind (Dialynas et al \cite{dia}). Stellar winds carry off a very negligible amount of mass from the stars while they provide (especially when magnetized) a very effective vehicle to provide angular momentum loss from the stars. The angular momentum loss is very crucial for a protostar because the protostar would otherwise breakup due to huge centrifugal forces at the equator that develop during the condensation process of the parental gas cloud. The extended active heating of the corona\footnote{This is believed to be caused by the dissipation of plasma waves produced by microflares in the coronal holes (Parker \cite{Park},\cite{Park1}). The microflares constitute magnetic reconnection produced via twisting of magnetic field lines by turbulent motion of their foot points on the stellar surface (Parker \cite{Park2}). In this paper, we will not address this issue because the stellar wind dynamics is contingent only on a heated corona but not so much on the details of the heating mechanism which are not yet clear (Klimchuk \cite{Kli}). It may be mentioned that there is however some belief that the problem of stellar wind acceleration should really be considered together with the problem of coronal heating (Holzer et al. \cite{Hol}), which is however a formidable effort.} (which provides for the high solar wind velocity observed at the orbit of Earth) coupled with high thermal conduction leads to expansion of the outer corona away to infinity, which seems to be adequate for weak to moderate stellar winds. However, the high speeds often observed in stellar winds do not appear to be achievable via control of conditions at the coronal base while thermal conduction does not appear to be capable of carrying sufficient energy into the corona to accelerate the wind speeds to $O$(100km/sec) or greater, in the case of the Sun (Parker \cite{Par}, \cite{Par1})\footnote{Thermal conduction from heat sources associated with dissipation of plasma waves at the coronal base could however generate tenuous supersonic winds (Parker, private communication, 2016).}. It is therefore believed that some additional acceleration mechanism must be operational beyond the coronal base, the understanding of which is one of the most challenging issues of stellar physics research. Parker \cite{Par} gave an ingenious stationary model for the stellar wind, which enables the stellar wind to accelerate smoothly through transonic speeds via a continuous conversion of the thermal energy in the wind into the kinetic energy of the outward flow. So, the stellar wind starts with subsonic speeds at the coronal base and expands continuously to supersonic speeds at large distances. In the case of the Sun, the existence of the solar wind was confirmed and its properties were revealed by {\it in situ} observations from planetary probes (like Mariner II (Neugebauer and Snyder \cite{neu}))satellites (Hundhausen \cite{Hun}, Meyer-Vernet \cite{Mey}).

\indent  The dragging out of a protostar's magnetic field by the stellar wind leads to an enhancement of the outward transfer of the angular momentum of the protostar and hence provides a mechanism for the braking of the protostar's rotation rate - {\it magnetic braking} (Schatzmann \cite{Sch}). The corona and the stellar wind are fully ionized so the magnetic field lines, which are anchored into the coronal base, remain {\it frozen} in the stellar wind as it streams outward, and enforce a {\it co-rotation}\footnote{In the region $r\lesssim r_A$, the strong magnetic field forces the wind to co-rotate with the star. However, in the region $r\gtrsim r_A$, with a weak magnetic field there is negligible magnetic torque to spin up the stellar wind; the angular momentum conservation in the wind then causes the azimuthal speed to decrease outward like $1/r$; as a result, the rotational effects become very weak farther away from the star. The weaker magnetic field frozen in the wind is swept by the wind and is forced to wrap up in a {\it spiral} pattern in the equatorial plane (Parker \cite{Par}). {\it In situ} observations by {\it Mariner II} confirmed that the average interplanetary magnetic field (IMF) showed this {\it spiral} structure (Ness and Wilcox \cite{NeWi}).} of the stellar wind as if it were a solid body out to the {\it Alfv\'{e}n radius} $r_A$ (where $v_r=B_r/\sqrt{\rho}$)\footnote{Numerical simulation (ud-Doula et al. \cite{ud}) with a dipole magnetic field showed that the Alfv\'{e}n radius $r_A$ is just a little beyond the maximum extent of closed magnetic loops in the stellar magnetosphere.} with a larger effective moment of inertia. This leads to an enhanced outward transfer of angular momentum when the stellar wind eventually escapes at great distances from the star, hence spinning the star down. This process indeed caused the Sun to lose most of its initial angular momentum (the equatorial rotation speed of the Sun is about 2km/sec) while the planets did not lose any.

\indent Near the surface of the star, the magnetization effects in the stellar wind are small and the stellar wind follows Parker's hydrodynamic model \cite{Par} reasonably well. However, Parker's hydrodynamic model \cite{Par} yields solar wind speeds at Earth's orbit which are much lower than the values typically observed. On the other hand, farther away, the magnetization effects become strong.\footnote{The magnetic pressure at $1 A U $ is comparable to the gas pressure in the solar wind, so the magnetization effects become as important as the thermal effects in the wind. Indeed, the existence of the solar wind was surmized in the 1950's as a result of the evidence that small uninterrupted variations in the earth's magnetic field ({\it geomagnetic} activity) were caused by observable phenomena on the Sun ({\it solar} activity) (example: {\it aurora borealis} (Birkeland \cite{Bir}), cometary tails (Biermann \cite{Bie})).}  Parker \cite{Par1} therefore suggested an additional acceleration mechanism in which energy is transferred to the solar wind by magnetohydrodynamic (MHD) waves propagating away from the Sun hence accelerating the solar wind to high speeds. Besides, the observed variability (spatial as well as temporal) in the solar wind has been traced to the presence of the MHD waves in the interplanetary medium (Marsch and Tu \cite{Mar}, Roberts and Goldstein \cite{Rob}, MacGregor and Charbonneau \cite{Char}, among others).  The MHD version of Parker's hydrodynamic model was considered by Modisette \cite{Mod}, Weber and Davis \cite{Weda} and Belcher and MacGregor \cite{Bema}. 

For moderately rotating stars, the stellar wind velocity needs to be considered both in the radial and azimuthal directions. In this paper, the effects of the stellar rotation and the consequent azimuthal stellar wind flow in Parker's \cite{Par} hydrodynamic stellar wind model are therefore discussed. Of special interest is the emergence of a whole new {\it hydrodynamic} physics via a new critical point in the stellar wind flow, which supersedes the critical point in Parker's \cite{Par} hydrodynamic model. The purpose of the present discussion is to put forward detailed considerations to show that the effect of the stellar rotation is to lead to tenuous and faster stellar wind flows, and hence to provide an efficient physical mechanism for protostars and strong rotators to lose their angular momentum quickly.
\vspace{.20in}

\noindent\large\textbf{2. Parker's Supersonic Stellar Wind Model}\\

Parker's hydrodynamic\footnote{See Parker \cite{Par1} for a justification of the fluid model, the tenuous state of the corona notwithstanding. A kinetic model is however imperative to deal with non-equilibrium particle velocity distributions and deviations from classical transport properties in space plasmas (Meyer-vernet \cite{Mey1}).} model \cite{Par} provided an ingenious development to show that a gas flow in a perpetual divergent channel (like the interplanetary space), thanks to the presence of a retarding body force (like gravity), can evolve from subsonic (at the coronal base) to supersonic (in the interplanetary space) speeds without requiring a physical throat section (as in a typical rocket nozzle situation). In this model, the stellar wind is represented by a steady, spherically symmetric\footnote{It may be pointed out that observations of the solar wind (Wang and Sheelay \cite{Wan}) suggested that, in reality, the stellar wind does not expand radially. Kopp and Holzer \cite{Kop} proposed a rapidly-diverging coronal flow tube which widens superradially (i.e., faster than $r^2$) in some active regions like the coronal hole (confirmed by Ulysses' polar pass observations (McComas et al. \cite{McC})). Such an arrangement has been found to lead to more tenuous and faster stellar winds.} flow so the flow variables are independent of time\footnote{In reality, the solar wind is found to exhibit a substantial time dependence, even at times of low solar activity.} and depend only on $r$, the distance from the star. The flow velocity is taken to be in the radial direction - either inward (accretion model) or outward (wind model). The equation of conservation of mass is 

\beq%1
\frac{2}{r}+\frac{1}{\rho}\frac{d\rho}{dr}+\frac{1}{v_r}\frac{dv_r}{dr}=0
\eq
	
\noindent standard notation being used.

Assuming the gravitational field to be produced by a central mass $M$, Euler's equation of momentum balance gives 

\beq%2
\rho v_r\frac{dv_r}{dr}=-\frac{dp}{dr}-\frac{GM}{r^2}\rho
\eq

\noindent $M$ being the mass of the star and $G$ being the gravitational constant.

The gas is assumed to be a perfect gas obeying an equation of state (which replaces the energy balance equation),

\beq%3
p=\rho RT
\eq
	
\noindent $R$ being the perfect gas constant.

For analytic simplicity, the flow is assumed to occur under isothermal ($T=constant$) conditions, so (3) becomes

\beq%4
p=a^2\rho
\eq

\noindent where $a$ is the constant speed of sound. For hot stars, isothermality is sustained by a strong radiation field. On the other hand, coronal heating (Gombosi et al. \cite{gom} and the high electron thermal conductivity tends to validate this approximation for a cooler solar-type corona, the cooling down of the solar wind while expanding through the interplanetary space notwithstanding. Indeed, SOHO observations (Cho et al. \cite{cho}) confirmed that the solar wind expands isothermally to considerable distances while spacecraft measurements have revealed (Hundhausen \cite{Hun}) that the plasma temperature drops only by a factor of 10 from the inner corona to the orbit of the earth\footnote{Parker \cite{Park3} showed that the isothermality assumption however turns out to be not very critical. (Parker \cite{Par2}) showed, in particular, that any temperature profile falling off more slowly than $1/r$ was shown to be compatible. Further, the temperature distribution only in the subcritical region ($r<r_\ast$, see below) was shown to be of relevance to the existence of a supersonic stellar wind. On the other hand, this implies that some energy input is crucial for sustaining the acceleration of the stellar wind (Lamers and Cassinelli \cite{Laca}). A more sound but less well-defined approach is to use the total energy balance equation that incorporates a reasonable set of physical assumptions about the mechanisms transporting energy in the corona and the stellar wind, and in particular, includes a source term in it which accounts for all the heating due to plasma waves, turbulence and thermal conductivity.}.

We assume again for analytic simplicity that the flow variables as well as their derivatives vary continuously so there are no shocks anywhere in the region under consideration.

Equations (1), (2), and (4) lead to\footnote{The structure of the stellar winds turns out to be similar for isothermal and polytropic gas flows (Maciel \cite{Mac}). If gas is assumed to exhibit instead a polytropic behavior, according to $p\sim\rho^\gamma$, equation (5) turns out to maintain its form so (6) still holds and the dynamics remains qualitatively unchanged. However, $a$ is no longer constant but varies with $r$, which leads to quantitative changes in the dynamics.} 

\beq\tag{5a}%5a
\lp v_r-\frac{a^2}{v_r}\rp\frac{dv_r}{dr}=\frac{2a^2}{r}-\frac{GM}{r^2}
\eq

\noindent or

\beq\tag{5b}%5b
\frac{1}{v_r}\frac{dv_r}{dr}=\frac{2a^2\lp r-r_*\rp}{r^2\lp v_r^2-a^2\rp}
\eq
	
\noindent where,

\beq\tag{6}%6
r_*\equiv\frac{GM}{2a^2}.
\eq

\noindent Observe that, at the critical point $r=r_\ast$, the gravitational energy of the stellar wind is comparable to its thermal energy, and the location of the critical point is strongly influenced by the extended coronal heating process.

Equation (5) exhibits a physically acceptable smooth solution, for which, we have at the \textit{critical point} $r=r_\ast$,

\beq\tag{7}
r=r_*:v_r=a
\eq

\noindent so the numerator and denominator in equation (5b) vanish simultaneously\footnote{In fact, if one assumes $r\approx r_\ast:v_r\approx a$, and hence puts

\beq\notag
y\equiv v_r-a,~ x\equiv r-r_*
\eq

\noindent the \textit{local} behavior of the solution, near $r=r_*$, is given by

\beq\notag
yy^\prime\approx\lp\frac{a^2}{r_*^2}\rp x.
\eq

\noindent This leads to 

\beq\notag
y\approx\underline{+}\lp\frac{a}{r_*}\rp x
\eq

\noindent or

\beq\notag
\lp\frac{v_r}{a}-1\rp\approx\underline{+}\lp\frac{r}{r_*}-1\rp
\eq

\noindent which describe the asymptotes near the point $v_r=a$, $r=r_\ast$ to the hyperbolas given by (8). Physically, the branch corresponding to the $+$ sign describes smooth acceleration of the flow through the transonic regime $(M\equiv v_r/a\underset{\sim}{<} 1 ~\text{to}~ M \underset{\sim}{>}1)$.}. 

The solution of equation (5), upon imposing the smoothness condition (7), is

\beq\tag{8a}
\lp\frac{v_r}{a}\rp^2-\log\lp\frac{v_r}{a}\rp^2=4 \log\lp\frac{r}{r_*}\rp+4\lp\frac{r_*}{r}\rp-3
\eq

\noindent or

\beq\tag{8b}
v_r e^{-\lp v_r^2/2a^2\rp}=a\lp\frac{r_*}{r}\rp^2 e^{\lp3/2-2r_*/r\rp}.
\eq

\noindent (8) exhibits two branches, for $r>r_*$,

\begin{itemize}
  \item [*] subsonic (Bondi's \cite{Bon} accretion problem);
  \item [*] supersonic (Parker's \cite{Par} wind problem).\end{itemize}

If $v_r<a$ at the critical point $r=r_\ast$, then according to equation (5b), $v_r$ attains its maximum at the critical point, and the gas flow remains subsonic everywhere. This is consistent with the belief that, for the case with the critical point high in the corona, most of the heating occurs in the subsonic low corona and the stellar wind turns out to be dense and slow - the {\it stellar breezes} (Leer and Holzer \cite{Lee}, Pneuman \cite{Pne}). On the other hand, if $v_r>a$ at the critical point, then according to equation (5b), $v_r$ attains its minimum at the critical point and the gas flow {\it chokes} so the sonic conditions shift to the critical point. This is consistent with the belief that, for the case with the critical point low in the corona, the heating (via the dissipation of plasma waves), which provides for the additional energization of the stellar wind, occurs mainly in the supersonic outer corona\footnote{Leer and Holzer \cite{Lee} showed that the observed high speeds in the solar wind is contingent on the energy being added to the gas flow after it has attained supersonice speed. By contrast, energy input in the inner corona increases the gas density and the mass flux but not the flow speed.} (as with the {\it afterburner} in an aircraft jet engine (Parker \cite{Par2})) and the stellar wind turns out to be tenuous and fast without changing the mass flux (Cranmer \cite{Cra1}). Thus, Parker's supersonic stellar wind solution (8) is indeed not {\it "finely-tuned"}, as sometimes misunderstood, in assuming the stellar wind to accelerate smoothly through the sonic conditions {\it exactly} at the critical point (Parker \cite{Par3}).

 For $r\ll r_*$, (8b) leads to 

\beq\tag{9}
r\ll r_*:v_r \approx a \lp\displaystyle\frac{r_*}{r}\rp^2 e^{\lp 3/2-2r_*/r\rp}
\eq

\noindent The corresponding density profile, on using equation (1) reexpressed as 

\beq\tag{10}
\rho r^2 v_r =const
\eq

\noindent is given by

\beq\tag{11}
\rho \sim e^{2r_\ast/r}
\eq

\noindent which is simply the one given earlier by Chapman \cite{Cha} using a hydrostatic model.  The neglect of the radial flow and the validity of the hydrostatic model in the subcritical region $r\ll r_*$ is due to the  corona strongly bound by the stellar gravity and governed by a near-hydrostatic force balance condition in this region (Cranmer \cite{Cra}). Lamers and Cassinelli \cite{Laca} indeed numerically demonstrated that the corona is almost in hydrostatic equilibrium not just at its base but until close to the critical point (6). The decrease of $\rho$, as per (11), faster than $1/r^2$, implies increase of $v_r$ with $r$,  the divergence of the interplanetary flow-geometry notwithstanding, to comply with equation (10). Thus, in the Parker  hydrodynamic model \cite{Par}, the flow acceleration is compressibility driven in the subsonic region. 

For $r\gg r_*$, (8a) indeed leads to 

\beq\tag{12}
r\gg r_*:v_r\approx 2a[\log\lp r/r_*\rp]^{1/2}
\eq

\noindent which implies that the radial outflow velocity increases slowly but indefinitely\footnote{SOHO observation have shown (Cho et al \cite{cho}) the solar wind is mainly accelerated below $10 r_\odot$ and then continues at a nearly constant speed.} This unphysical result may be traced to the assumption of isothermality of the wind (which implies an extended coronal heating and is contingent on the continuous addition of energy in the interplanetary space and would not be tenable at large distances from the star). 

On the other hand, using (12), equation (10) shows that 

\beq\tag{13}
\rho\sim 1/r^2
\eq

\noindent so the density falls off algebraically, like $1/r^2$, at large distances from the star. 

\vspace{.20in}

\noindent\large\textbf{3. Effective de Laval Nozzle in the Parker Hydrodynamic Model}\\

The continuous acceleration of the stellar wind flow, as per the Parker hydrodynamic model \cite{Par}, from subsonic speeds at the coronal base to supersonic speeds in the interplanetary space immediately leads one to suspect a \textit{de Laval} type nozzle mechanism (Clauser \cite{Cla}, Parker \cite{Par4}) implicit in the Parker hydrodynamic model \cite{Par}. (Notwithstanding the conceptual facilitation rendered by the de Laval nozzle analogy in this context, it should be pointed out that this analogy was never intended to provide a framework for serious calculation (Parker, private communication, 2016)). Indeed, on a superficial level, one may visualize an effective de Laval type nozzle associated with the Parker hydrodynamic model \cite{Par} and determine its geometry.

For a one-dimensional gas flow in a de Laval nozzle with cross section area $A=A\lp x\rp$, we have (Shivamoggi \cite{Shi}),

\beq\tag{14}
\frac{1}{v}\lp 1-\frac{v^2}{a^2}\rp\frac{dv}{dx}=-\frac{1}{A}\frac{dA}{dx}.
\eq

If $\tilde A =\tilde A \lp r\rp$ is the cross section area of an effective de Laval nozzle associated with the Parker hydrodynamic model \cite{Par}, we have from equation (5a), 

\beq\tag{15}
\frac{1}{v_r}\lp 1-\frac{v_r^2}{a^2}\rp \frac{dv_r}{dr}=-\frac{2}{r^2}\lp r-r_*\rp=-\frac{1}{\tilde A}\frac{d\tilde A}{dr}
\eq

\noindent from which, we obtain,

\beq\tag{16}
\tilde A\lp r\rp=4\pi r^2\left[e^{-2r_*/r_0\lp 1-r_0/r\rp}\right].
\eq

\noindent Comparison of (16) with (11) shows that the multiplicative correction to the divergent channel needed to yield the effective de Laval nozzle associated with the Parker hydrodynamic model \cite{Par} corresponds just to the Chapman \cite{Cha} hydrostatic density profile, which encapsulates the dominant effects of gravity near the star.

It is also easy to verify that the effective de Laval nozzle cross section area $\tilde A \lp r\rp$ shows a physical throat section at $r=r_*$, as to be expected. In fact,

\beq\tag{17a}
r\lessgtr r_*:\tilde A^\prime\lp r\rp=8\pi\lp r-r_*\rp  e^{-2r_*/r_0\lp 1-r_0/r\rp}\lessgtr 0
\eq

\noindent and further,

\beq\tag{17b}
r\approx r_*:\tilde A \lp r\rp \approx 4\pi e^{2\lp 1-r_*/r_0\rp}\left[r^2_*+\lp\triangle r\rp^2\right],\triangle r\equiv r-r_*.
\eq

\noindent So, $\tilde A \lp r\rp$ has a minimum at $r=r_*$, like a de Laval nozzle!

However, the effective nozzle associated with the Parker hydrodynamic model \cite{Par} turns out to mimic de Laval, only in a superficial way and discrepancies arise with regard to the density variation aspect. This may be seen by noting that we have for a one-dimensional gas flow in a de Laval nozzle (Shivamoggi \cite{Shi}),

\beq\tag{18}
\frac{1}{\rho}\frac{d\rho}{dx}=-\frac{v^2}{a^2}\frac{1}{v}\lp \frac{dv}{dx}\rp < 0,~ \forall x
\eq

\noindent which implies, 

\beq\tag{19}
\begin{matrix}
\begin{aligned}
&*v\ll a: \displaystyle\frac{d\rho}{dx} ~ \text {small}\\
\\
&*v\gg a: \displaystyle\frac{d\rho}{dx}~\text {large }. 
\end{aligned}
\end{matrix}
\eq

\noindent On the other hand, we have for Parker's hydrodynamic model \cite{Par}, from equation (5b), 

\beq\tag{20}
\begin{matrix}
\begin{aligned}
&r\ll r_*:\displaystyle \frac{1}{v_r}\displaystyle\frac{dv_r}{dr}\approx 2 \displaystyle\frac{r_*}{r^2}\\
\\
&r\gg r_*: \displaystyle\frac{1}{v_r}\displaystyle\frac{dv_r}{dr}\approx 2 \lp\displaystyle\frac{a^2}{v_r^2}\rp\frac{1}{r}.
\end{aligned}
\end{matrix}
\eq

\noindent Using (20), we have, from equations (1) and (12),

\beq\tag{21}
\frac{1}{\rho}\frac{d\rho}{dr}\approx
\left\{
\begin{matrix}
\begin{aligned}
&\displaystyle-\frac{2}{r}\lp 1+\frac{r_*}{r}\rp \approx -2\frac{r_*}{r^2},~ r\ll r_*\\
\\
&-\displaystyle\frac{2}{r}\lp 1+\frac{a^2}{v_r^2}\rp \approx -\frac{2}{r}\left[1+\frac{1}{4\log\lp r/r_*\rp}\right],~r\gg r_*
\end{aligned}
\end{matrix}\right.
\eq

\noindent which implies

\beq\tag{22}
\begin{matrix}
\begin{aligned}
&*r\ll r_*: \frac{d\rho}{dr}\text {~large }\\
\\
&*r\gg r_*: \frac{d\rho}{dr} \text {~small}
\end{aligned}
\end{matrix}\eq

Comparison of (19) and (22) shows that the fluid density drops drastically in the subcritical region and essentially remains constant in the supercritical region in Parker's hydrodynamic model \cite{Par}, while the opposite is true for a de Laval nozzle. More specifically, in the de Laval nozzle, the flow acceleration is flow-geometry driven in the subsonic region while it is compressibility driven in the supersonic region, while the opposite is true for Parker's hydrodynamic model \cite{Par}.

\vspace{0.15in}

\noindent\large\textbf{4. Stellar Wind with Azimuthal Flow: Hydrodynamic Model}\\

For moderately rotating stars, the stellar wind velocity needs to be considered both in the radial and azimuthal directions\footnote{Though for the current Sun, the solar wind, thanks to the high thermal conductivity, seems to be mostly thermally driven, the solar wind was primarily centrifugal driven for the rapidly rotating young Sun (Schrijver \cite{sch}), as with any protostar.}. Hartman and MacGregor \cite{Hart} gave a full-fledged MHD treatment for this problem. We give here a much simpler hydrodynamic treatment which indeed turns out to be adequate for this aspect. The radial component of the equation of momentum balance is then given by

\beq\tag{23}
\rho\lp v_r\frac{dv_r}{dr}-\frac{v_\phi^2}{r}\rp =-\frac{dp}{dr}-\frac{\rho GM}{r^2}.
\eq

Thanks to the co-rotation of the stellar wind enforced by the stellar magnetic field, the stellar wind exhibits a rigid-body rotation out to the Alfv\'{e}n radius $r\approx r_A$ (see also Section 5), so we may take 

\beq\tag{24}
v_\phi\approx\Omega_\ast r
\eq

\noindent where $\Omega_\ast$ is the angular velocity of the star. Using (3) and (4) further, equation (23) leads to 

\beq\tag{25a}
\lp v_r-\frac{a^2}{v_r}\rp \frac{dv_r}{dr}=\frac{2a^2}{r^2}\lp r-r_\ast+\frac{\Omega_\ast^2 r^3}{2a^2}\rp.
\eq

It may be mentioned that equation (25a) was also deduced by Hartmann and MacGregor \cite{Hart} from their full-fledged MHD treatment\footnote{It may be placed on record that the author became aware of the work of Hartmann and MacGregor \cite{Hart} only after the present work was completed.} with the apparent implication that it is an exclusive MHD result. This development obscures the essential hydrodynamic physics underlying equation (25a). The present formulation reveals that the latter is a full-fledged hydrodynamic result barring the interjection of the stellar-wind co-rotation ansatz. We will now discuss the details of the stellar wind physics described by equation (25a), not given before. 

\vspace{.20in}

\noindent (i) {\bf New Critical Point}\\

Equation (25a) may be rewritten as

\beq\tag{25b}
\frac{1}{v_r}\frac{dv_r}{d\xi}=\displaystyle\frac{\Omega_\ast^2 \tilde{r}^2\lp\xi^3+\gamma\xi-1\rp}{\xi^2\lp v_r^2-a^2\rp}
\eq

\noindent or

\beq\tag{25c}
\frac{1}{v_r}\frac{dv_r}{d\xi}=\frac{\Omega_\ast^2 \tilde{r}^2\lb\xi-\lp\alpha-\beta\rp\rb\lb\lp\xi+\displaystyle\frac{\alpha-\beta}{2}\rp^2+\displaystyle\frac{3}{4}\lp\alpha+\beta\rp^2\rb}{\xi^2\lp v_r^2-a^2\rp}
\eq

\noindent where, 

\beq\tag{25d}
\begin{matrix}
\xi\equiv\displaystyle\frac{r}{\tilde{r}}, \tilde{r}\equiv\lp\displaystyle\frac{GM}{\Omega_\ast^2}\rp^{1/3}\\
\alpha, \beta\equiv\lb\displaystyle\sqrt{\displaystyle\frac{1}{4}+\frac{\gamma^3}{27}}~\underline{+}~\frac{1}{2}\rb^{1/3},~ \gamma\equiv\displaystyle\frac{\tilde{r}}{r_\ast}=2\lp\frac{a^3}{GM\Omega_\ast}\rp^{2/3}.
\end{matrix}
\eq

Equation (25c) shows that the critical point in a stellar wind with azimuthal flow is given by\footnote{The cubic equation, $$\xi^3+\gamma\xi-1=0$$ has one real root $\lp\alpha-\beta\rp$ and two complex conjugate roots (as seen from (25c)).}

\beq\tag{26a}
r=\hat{r}\equiv\lp\alpha-\beta\rp\tilde{r}:v_r=a
\eq

\noindent where the numerator and denominator in equation (25b) vanish simultaneously, so equation (25a) exhibits a physically acceptable smooth solution.

For strong rotators, $\gamma <<1$, and (26a) leads to

\beq\tag{26b}
\hat{r}\approx\lp1-\frac{\tilde{r}}{3r_\ast}\rp\tilde{r}\approx\tilde{r},~\frac{\tilde{r}}{r_\ast}<<1.
\eq

\indent So, for strong rotators the critical point (26a) is essentially determined by the basic stellar parameters like the mass $M$ and the angular velocity $\Omega_\ast$ and supersedes the critical point (6) (that is determined by the speed of sound $a$ in the stellar wind) in Parker's hydrodynamic model \cite{Par}. Observe that at the critical point $r\approx\tilde{r}$, the corotation speed $\Omega_\ast \tilde{r}$ of a particle is equal to the circular {\it Keplerian orbit} speed $\sqrt{GM/\tilde{r}}$ This also corresponds to the balance between outward centrifugal force and the stellar gravity force. So, Keplerian-orbit conditions can be expected to prevail near strong rotators. 

For slow rotators $\gamma=\lp a^3/GM\Omega_\ast\rp^{2/3}>>1,$ and (26a) leads to 

\beq\tag{26c}
\hat{r}\approx\lb 1-\frac{\sqrt{3}}{2}\lp\frac{r_\ast}{\tilde{r}}\rp^{3/2}\rb r_\ast,~ \frac{\tilde{r}}{r_\ast}>>1.
\eq

For the Sun, which is a moderate rotator, ($\Omega_\ast\approx 2\cdot8\times 10^{-6}rad/sec$), $\gamma\approx 1\cdot 8$, and (26a) gives\footnote{If $\lp\displaystyle\frac{\gamma}{3}\rp << 1$, (26a) leads to $$r\approx\lb 1-\lp\frac{\gamma}{3}\rp^{1/3}+O\lp\frac{\gamma}{3}\rp^{2/3}\rb\tilde{r}$$.}

\beq\tag{26c}
\hat{r}\approx \cdot 49 \tilde{r}\approx \cdot 87 r_\ast.
\eq

\noindent So, the critical point (26a) for the Sun, upon including the azimuthal flow in the solar wind, occurs lower in the corona than the critical point in Parker's hydrodynamic model \cite{Par}\footnote{A similar situation is found to prevail for superradial coronal flows like those near coronal holes (Parker \cite{Par4}).}, and in effect, signifies the conversion of rotational energy of the star into streaming energy of the stellar wind.  This shift in the critical point becomes even larger for strong rotators, for which, $\tilde{r}<r_\ast$. As a consequence of this, the stellar wind also experiences a stronger \textit{afterburner} action in the corona and turns out to be tenuous and fast (Section 2). Besides, unlike the critical point in Parker's hydrodynamic model which lies at different heights above the stellar surface in different stellar wind environments (like above solar polar caps and solar streamers), the critical point $r\approx\tilde{r}$  for a strong rotator seems to be at a fixed location for a given star.

\vspace{.20in}

\noindent (ii) {\bf The Analytical Solution and Asymptotic Results}\\

The solution of equation (25), upon imposing the flow smoothness condition (26a), is 

\beq\tag{27a}
\frac{v_r^2}{a^2}- \log \lp\frac{v_r^2}{a^2}\rp=4\log\lp\frac{r}{\hat{r}}\rp+4\frac{r_\ast}{r}+\frac{\Omega_\ast^2r^2}{a^2}-\lp 4\frac{\hat{r}_\ast}{\hat{r}}+\frac{\Omega_\ast^2\hat{r}^2}{a^2}-1\rp
\eq

\noindent or

\beq\tag{27b}
\displaystyle v_r e^{-\lp v_r^2/2a^2\rp}=a\lp\frac{\hat{r}}{r}\rp^2 \displaystyle e^{\displaystyle\lb\frac{2r_\ast}{\hat{r}}\lp 1-\frac{\hat{r}}{r}\rp + \frac{\Omega_\ast^2\hat{r}^2}{2a^2}\lp 1-\frac{r^2}{\hat{r}^2}\rp -\frac{1}{2}\rb}
\eq

For $r<<\hat{r}$, (27b) leads to 

\beq\tag{28}
\displaystyle v_r\approx a \lp\frac{\hat{r}}{r}\rp^2\displaystyle e^{\displaystyle\lb\frac{2r_\ast}{\hat{r}}\lp 1-\frac{\hat{r}}{r}\rp + \frac{1}{2}\lp\frac{\Omega_\ast^2\hat{r}^2}{a^2}-1\rp\rb}
\eq

\noindent The corresponding density profile, on using equation (10), is given by

\beq\tag{29}
\rho\sim e^{2r_\ast/r}
\eq

\noindent (29) is again the one given earlier by Chapman \cite{Cha} using a hydrostatic model, which is palusible because the stellar rotation effects are negligible in this region.

For $r>>\hat{r}$, (27a) leads to 

\beq\tag{30}
v_r\approx\Omega_\ast r.
\eq

\noindent Comparison  of (30) with (12) shows that the effect of stellar rotation is to cause enhanced acceleration of the stellar wind at large distances from the star.

The density profile corresponding to (30), on using (10), is given by

\beq\tag{31}
\rho\sim\frac{1}{r^3}.
\eq

\noindent So, at large distances from the star, comparison  of (31) with (13) shows that the stellar rotation causes the density to fall off faster, like $1/r^3$, than the non-rotating star result (13).

\vspace{.20in}

\noindent (iii) {\bf Strong Rotators}\\

For strong rotators $\lp\gamma<<1\rp$, equation (25a) becomes 

\beq\tag{32}
\lp v_r-\frac{a^2}{v_r}\rp \frac{dv_r}{dr}\approx \Omega^2_\ast\lp r-\frac{\tilde{r}^3}{r^2}\rp
\eq\\

\noindent from which, upon imposing the flow smoothness condition (26a), 

\beq\tag{33a}
\frac{v_r^2}{a^2}-\log \lp\frac{v_r^2}{a^2}\rp\approx\frac{\Omega^2_\ast}{a^2}\lp r^2+\frac{2\tilde{r}^3}{r}-3\tilde{r}^2\rp
\eq

\noindent or

\beq\tag{33b}
\displaystyle v_r e^{-\lp v_r^2/2a^2\rp}\approx \displaystyle e^{\displaystyle-\frac{\Omega^2_\ast}{2a^2}\lp r^2+\frac{2\tilde{r}^3}{r}-3\tilde{r}^2\rp}.
\eq

For $r<<\tilde{r}$, (33b) leads to 

\beq\tag{34}
v_r\approx ae^{-\lp 2r_\ast/r\rp}.
\eq

\noindent Comparison of (34) with (28) shows that the effect of stellar rotation is to cause enhanced acceleration of the stellar wind even near the star.

The density profile corresponding to (34), on using equation (10), is given by

\beq\tag{35}
\rho\sim\frac{1}{r^2} e^{2r_\ast/r}
\eq

\noindent which shows that, for the strong rotator case, the density, even close to the star, falls off faster than (29) given by Chapman's \cite{Cha} hydrostatic model.

For $r>>\tilde{r}$, equation (32) leads to 

\beq\tag{36}
v_r\approx \Omega_\ast r.
\eq

\noindent Comparison of (36) with (30) shows that the stellar wind flow, at large distances from the star, is qualitatively insensitive to the magnitude of the stellar rotation.

\vspace{.20in} 

\noindent (iv) {\bf Effective de Laval Nozzle}\\

If $\tilde{A}=\tilde{A}\lp r\rp$ is the cross section area of an effective de Laval nozzle associated with the stellar wind flow, we have from equation (23), 

\beq\tag{37}
\frac{1}{v_r}\lp1-\frac{v_r^2}{a^2}\rp\frac{dv_r}{dr}=-\frac{2}{r^2}\lp r-r_\ast-\frac{\Omega^2_\ast r^3}{2a^2}\rp=-\frac{1}{\tilde{A}}\frac{d\tilde{A}}{dr}
\eq

\noindent from which, we obtain

\beq\tag{38}
\tilde{A}\displaystyle\lp r\rp=4\pi r^2\displaystyle e^{\displaystyle \lb-2r_\ast/r_o \lp 1-r_o/r\rp+\Omega^2_\ast r^2_o/2a^2 \lp r^2/r^2_o-1\rp\rb}.
\eq

It is easy to verify that the effective de Laval nozzle cross section area $\tilde{A}\lp r\rp$, given by (38), shows again a physical throat section at $r=\hat{r}$, as expected. In fact, 

\beq\tag{39}
r\lessgtr\hat{r}:\tilde{A}\lp r\rp=\frac{8\pi\Omega^2_\ast}{a^2}\lp r^3+\frac{2a^2}{\Omega^2_\ast}r-\tilde{r}^3\rp e^{\lb-2r_\ast/r_o \lp 1-r_o/r\rp+\Omega_\ast^2 r^2_o/2a^2\lp r^2/r^2_o-1\rp\rb}\lessgtr 0
\eq

\noindent and further, 

\beq\tag{40}
\begin{matrix}
r\approx\hat{r}:\tilde{A}\lp r\rp \approx 4 \pi e^{\lb-2r_\ast/\hat{r}r_o + \Omega^2_\ast/2a^2     \lp \hat{r}+r_o\rp\rb \lp\hat{r}-r_o\rp}\\
\\
\times \lb \hat{r}^2+\lp \displaystyle\frac{3\Omega_\ast^2\hat{r}^2}{a^2}+1\rp \lp\Delta r\rp^2\rb, \Delta r\equiv r-\hat{r}.
\end{matrix}
\eq

\vspace{.20in}

\noindent So, $\tilde{A}\lp r\rp$ has a minimum at $r=\hat{r}$, for a stellar wind with azimuthal flow, like a de Laval nozzle. It is of interest to note that the stellar rotation causes the physical throat of the effective de Lavel nozzle to become narrower, which underscores the enhanced acceleration of the stellar wind.

\vspace{0.15in}
\noindent\large\textbf{5. Discussion}\\

Following Parker's \cite{Par} ingenious demonstration that a stellar wind, thanks to a retarding body force (like gravity), can evolve from subsonic to supersonic speeds (as confirmed by {\it in situ} observations from satellites) even in a purely divergent channel (like the interplanetary space), Parker's supersonic stellar wind model \cite{Par} has been a topic of enormous research\footnote{It may be mentioned that while Parker's hydrodynamic  model \cite{Par} is known to provide an excellent first-order approximation (Schrijver \cite{sch}), there are some observed aspects of the stellar wind that are not fully explained by Parker's hydrodynamic model \cite{Par}. An example is the fast stellar wind (Feldman et al. \cite{Fel}), which is channeled out of the coronal holes associated with funnel-like expansion of radial open magnetic field lines of largely single polarity (as revealed by SOHO observations, Hassler et al. \cite{Has}) and, despite the cooler state of the coronal holes, accelerates at a pace considerably fast, at variance with that predicted by Parker's hydrodynamic model \cite{Par}.}. 

\noindent For moderately rotating stars, the stellar wind velocity needs to be considered both in the radial and azimuthal directions. In this paper, we have therefore made detailed considerations of the effects of the stellar rotation and the consequent azimuthal stellar wind flow in Parker's \cite{Par} hydrodynamic stellar wind model. Of special interest is the emergence of a whole new hydrodynamic physics via a new critical point in the stellar wind flow, which supersedes the critical point in Parker's \cite{Par} hydrodynamic model. The effect of the stellar rotation is shown to cause the new critical point to occur lower in the corona, so the stellar wind experiences a stronger \textit{afterburner} (as in an aircraft jet engine) action in the corona. For strong rotators, the new critical point, is shown to occur at a fixed location for a given star, determined only by the basic stellar parameters like the mass $M$ and the angular velocity $\Omega_\ast$, the variations in the stellar wind environment notwithstanding. The stellar rotation leads to stronger density fall-off and enhanced acceleration of the  stellar wind at large distances from the star - this effect materializes even close to the star, for strong rotators. The stellar rotation causes the physical throat section of the effective \textit{de Laval} nozzle associated with the stellar wind flow to become narrower, indicative of an enhanced flow acceleration. Thus, the stellar rotation leads to tenuous and faster stellar wind flows without change in the mass flux, and hence provides an efficient physical mechanism for protostars and strong rotators to lose their angular momentum quickly. It may be pointed out that the extended active heating of the corona may be represented in a first approximation by using the polytropic gas relation $p\sim\rho^\alpha$, where $\alpha <5/3$, the value corresponding to an adiabatic process (Parker \cite{Par5}). The generalization of the present formulatons and results for polytropic gas flows will be reported separately. Next, the stability of the steady solar wind solutions discussed above is an important question and some attempts, particularly on the hydrodynamic model, have been made (Jockers \cite{Joc}, Velli \cite{Vel}) and Parker's \cite{Par} supersonic stellar wind solution appears to be the only stable solution among all other stellar wind solutions (Velli \cite{Vel})\footnote{The spurions speculation that Parker's \cite{Par} supersonic stellar wind solution is "finely-tuned" had actually led to the false conjecture that Parker's supersonic stellar wind solution might be dynamically unstable (Parker \cite{Par6}).}. Finally, it may be noted that the distances of the stellar objects and the difficulty in accessing even the solar coronal base conditions render the observation at this time insufficient to fully corroborate a particular proposed acceleration mechanism for the stellar wind. The recent deployment of the Parker Solar Probe in the solar corona is expected to lead to considerable progress in this direction.
\vspace{0.15in}

\noindent\large\textbf{Acknowledgements}\\

This work was initiated during my stay at the Institute for Fusion Studies, University of Texas, Austin. I am thankful to Professor Swadesh Mahajan for his hospitality and helpful discussions. Part of this work was carried out when I held a visiting research appointment at the Eindhoven University of Technology, supported by a grant from The Netherlands Organization for Scientific Research (NWO). I am thankful to Professor Gert Jan van Heijst for his hospitality and helpful discussions. I am thankful to Professor Eugene Parker for his valuable advice and suggestions. I am thankful to Professors Grisha Falkovich, Katepalli Sreenivasan and Drs. Frank Primini and Peter Weichman for their helpful remarks. I am also thankful to Professor Michael Johnson for helpful discussions.

\end{document}